\documentclass{article}
\usepackage{spconf,amsmath,graphicx}
\usepackage{amsfonts}
\usepackage[ruled,vlined]{algorithm2e}
\usepackage{booktabs}
\usepackage[flushleft]{threeparttable}
\SetKwComment{Comment}{$\triangleright$\ }{}

\newtheorem{theorem}{Theorem}

\DeclareMathOperator{\tr}{tr}
\newcommand{\indep}{\perp \!\!\! \perp}
\DeclareMathOperator*{\argmax}{argmax}

\title{Modeling Dependent Structure for Utterances in ASR Evaluation}

\name{Zhe Liu, Fuchun Peng}
\address{Meta AI, Menlo Park, CA, USA}

\begin{document}
\ninept
\maketitle

\begin{abstract}
The \emph{bootstrap} resampling method has been popular for performing significance analysis on word error rate (WER) in automatic speech recognition (ASR) evaluation. To deal with dependent speech data, the \emph{blockwise bootstrap} approach is also introduced. By dividing utterances into uncorrelated blocks, this approach resamples these blocks instead of original data. However, it is typically nontrivial to uncover the dependent structure among utterances and identify the blocks, which might lead to subjective conclusions in statistical testing. In this paper, we present \emph{graphical lasso} based methods to explicitly model such dependency and estimate uncorrelated blocks of utterances in a rigorous way, after which blockwise bootstrap is applied on top of the inferred blocks. We show the resulting variance estimator of WER in ASR evaluation is statistically consistent under mild conditions. We also demonstrate the validity of proposed approach on LibriSpeech dataset.
\end{abstract}


\section{Introduction}
Word error rate (WER) is a common metric for the performance of an automatic speech recognition (ASR) system. Derived from the Levenshtein distance \cite{navarro2001guided}, WER can be calculated as
\begin{equation}
\label{wer}
    W:=\frac{\sum_{i=1}^n e_i}{\sum_{i=1}^n m_i},
\end{equation}
where $m_i$ is the number of words in the $i$th reference, and $e_i$ refers to the sum of insertion, deletion, and substitution errors computed from the dynamic string alignment of the recognized sequence with the reference sequence.

When it comes to compare the transcription quality of two ASR systems on the same evaluation dataset, the absolute or relative WER difference of system \emph{B} versus system \emph{A} is widely used:
\begin{align}
    \Delta W_{abs}:&={W}_B - {W}_A=\frac{\sum_{i=1}^n (e_i^B - e_i^A)}{\sum_{i=1}^n m_i}, \\
    \Delta W_{rel}:&=\frac{{W}_B-{W}_A}{{W}_A}=\frac{\sum_{i=1}^n (e_i^B - e_i^A)}{\sum_{i=1}^n e_i^A}.
\end{align}

Oftentimes, it is crucial to understand the reliability of computed WER on some evaluation dataset as well as the comparison of WERs between two ASR systems. For instance, we would like to answer if a WER of 0.05 on only 500 evaluation utterances is trustworthy on measuring the overall ASR quality for the target domain; we also want to know if a 2\% relative WER improvement comparing ASR system \emph{B} with ASR system \emph{A} is due to chance or real.

For these purposes, statistical testing on WER has been explored \cite{gillick1989some, pallett1993benchmark, strik2000comparing, bisani2004bootstrap, vilar2007human, vilar2008efficient, liu2020statistical}. In particular, the work of \cite{bisani2004bootstrap} presents a \emph{bootstrap} approach for significance analysis on ASR evaluation which makes no distributional approximations and is easy to use.

Specifically, given any sequence of independent and identically distributed random variables $\{U_i\}_{i=1}^n$, we would like to estimate the variance of some statistic $T(U_1, ..., U_n)$. The bootstrap method \cite{efron1994introduction, efron2003second} resamples data from the empirical distribution of $\{U_i\}_{i=1}^n$ for $B^{boot}$ times, and then re-calculates the statistic $T$ on each of these ``bootstrap'' samples. The variance of $T$ can be estimated from the sample variance of these computed statistics. In significance analysis for WER, $U_i$ represents the information of the $i$th utterance, and $T$ refers to the statistic of $W$ as in (\ref{wer}).

However, the vanilla bootstrap fails in dealing with dependent data, which is prevalent in the speech world. More recently, authors in \cite{liu2020statistical} propose the \emph{blockwise bootstrap} approach that addresses this issue by dividing utterances into nonoverlapping and uncorrelated blocks, then resampling these blocks instead of original data.

One key issue confronting the blockwise bootstrap estimator in \cite{liu2020statistical} is how to determine the block structure over utterances in ASR evaluation such that speech recognition errors in different blocks are uncorrelated. In general, speech recognition errors are considered as dependent if the corresponding utterances share similar acoustic or semantic characteristics. However, it is nontrivial to properly define such dependency in a rigorous way, which could lead to subjective conclusions and thus makes this approach less attractive in practical applications. A more advanced data-driven method is called for to quantify the dependent structure of utterances in ASR evaluation.

In this work, we propose to leverage \emph{graphical modeling} \cite{lauritzen1996graphical} to estimate the dependent structure among speech utterances in ASR evaluation. Undirected graphical models have become as a powerful tool of representing complex interactions among high-dimensional random variables. In our applications, each utterance is represented as an embedding vector with its acoustic and semantic information being encoded, and is considered as a random variable in the graph network. The inferred uncorrelated blocks of utterances, will be then used in the blockwise bootstrap estimator for performing statistical testing on WER.

Under the assumption of multivariate Gaussian among utterance representation random variables, the problem of dependent structure estimation is reduced to learning a sparse precision matrix and identifying its zero-pattern. The \emph{graphical lasso} \cite{yuan2007model, friedman2008sparse}, a sparse penalized maximum likelihood estimator for the precision matrix, has been  studied in recent years. We introduce this method and its variant on relaxing the Gaussian assumption \cite{liu2009nonparanormal} to the applications of significance analysis on ASR evaluations. 

Conventionally, clustering based methods can be used on utterance embeddings and partition them into groups. However, there is typically no theories to establish the uncorrelatedness between utterances from different clusters. Compared with these methods, the use of graphical modeling provides more statistical guarantees in the dependent structure inference for utterances. Moreover, our approach is able to estimate the entire conditional independence graph and corresponding correlation matrix, which is more flexible to use.

We mainly pursue three goals in this paper: (1) to present new estimation approaches for utterances' dependent structure with the use of graphical lasso and its variant, and illustrate their applications in significance analysis on ASR quality evaluations; (2) to provide theories on the consistency of the resulting variance estimator; (3) to demonstrate empirical results on LibriSpeech speech dataset using our proposed methods. To the best of our knowledge, our paper is the first to introduce graphical modeling based approaches in estimating underlying dependent structure among evaluation utterances and explicitly addressing dependent speech data in ASR evaluations. Besides the applications to WER significance analysis, the proposed methods could also be useful in error analysis for ASR, correlation-aware speech model training, among others.


\section{Methods}
\label{methodology}
Suppose $\{U_i\}_{i=1}^n$ is an evaluation dataset for ASR, consisting of $n$ utterances with the $i$th utterance denoted by $U_i$. Two ASR systems \emph{A} and \emph{B} are evaluated on $\{U_i\}_{i=1}^n$. Assume we have their evaluation results as follows:
\begin{equation}
    \{(m_i, e_i^A, e_i^B)\}_{i=1}^n,
\end{equation}
where $m_i$ is the number of reference words of the $i$th utterance, $e_i^A$ and $e_i^B$ represent the numbers of word errors in ASR systems \emph{A} and \emph{B}, respectively. The statistics of our interest are the WER of ASR system \emph{A} denoted as $W_A$, the absolute WER difference $\Delta W_{abs}$ and relative WER difference $\Delta W_{rel}$ comparing ASR system \emph{B} versus ASR system \emph{A}. In particular, we would want to compute the 95\% confidence intervals for these statistics.

\subsection{Graph Structure Estimation}
We treat $U_i$ as a random variable representing the $i$th utterance in the evaluation set. Denote $u_i$ as the $L$-dimensional embedding vector for $U_i$ with $u_i=(u_{i1}, \ldots, u_{iL})^T$. Let $\bar{u}_i:=\frac{1}{L}\sum_{l=1}^L u_{il}$ be the sample mean. In practice, the representation $u_i$ for the utterance $U_i$ can be obtained from the speaker embedding vector, or sentence embedding vector of the reference, or their concatenation.

Suppose the random vector $(U_1, \ldots, U_n)^T$ for utterances has the density $P^*$. Then an undirected graph $G$ for $P^*$ has $n$ vertices, collected in a set $V$, one for each variable. We represent the edges as a set $E$ of unordered pairs: $(i,j)\in E$ if and only if there is an edge between $U_i$ and $U_j$. In a conditional independence graph, an edge between $U_i$ and $U_j$ is absent if $U_i$ and $U_j$ are independent (denoted as $U_i \indep U_j$), given the other variables. With the observations of $\{u_i\}_{i=1}^n$, the goal is to estimate the graph structure $G$.

Under the assumption that $(U_1, \ldots, U_n)^T\sim\mathcal{N}_n(\mu, \Sigma)$, it has been shown that there is no edge between $U_i$ and $U_j$ if and only if $\Theta_{ij} = 0$, where $\Theta=\Sigma^{-1}$ is the precision matrix. Therefore, the graph estimation problem in Gaussian graphical models are equivalent to the estimation of a sparse precision matrix $\Theta$.

The graphical lasso estimator \cite{yuan2007model, friedman2008sparse} encourages some of the entries of the estimated precision matrix to be zero. Specifically, it maximizes the $\ell_1$-regularized log-likelihood  under the constraint that $\Theta$ is positive definite (denoted as $\Theta\succ0$):
\begin{align}
\label{glasso}
\widehat{\Theta}_\lambda:=\argmax_{\Theta\succ0}\biggl\{\log\det(\Theta)-\tr(\widehat{\Sigma}\,\Theta)-\lambda\sum_{i\neq j}|\Theta_{ij}|\biggr\},
\end{align}
where $\log\det(\cdot)$ denotes the log-determinant of a matrix, $\tr(\cdot)$ represents the trace of a matrix, $\lambda$ is a regularization parameter, and $\widehat\Sigma$ is the empirical covariance matrix:
\begin{align}
\widehat\Sigma_{ij}=\frac{1}{L - 1}\sum_{l=1}^L (u_{il}-\bar{u}_{i})(u_{jl}-\bar{u}_j).
\end{align}
The optimization problem in (\ref{glasso}) can be solved using the coordinate descent algorithm \cite{friedman2007pathwise, wu2008coordinate}. The regularization parameter $\lambda$ can be chosen via cross-validation in practical applications.

While the graphical lasso is useful in many scenarios, a reliance on exact normality can be limiting. The \emph{nonparanormal} \cite{liu2009nonparanormal}, a form of Gaussian copula, weakens the Gaussian assumption by imposing normality on the transformed random vector $(f_1(U_1), . . . , f_n(U_n))^T$, with each $f_i(\cdot)$ being a function that meets certain conditions. This allows arbitrary single variable marginal probability distribution in the model. 

Specifically, we say $(U_1, \ldots, U_n)^T$ follows a nonparanormal distribution if  there exist monotonic and differentiable functions $\{f_i(\cdot)\}_{i=1}^n$ such that $(f_1(U_1), . . . , f_n(U_n))^T\sim\mathcal{N}_n(\mu, \Sigma)$. Under this assumption, it was shown in \cite{liu2009nonparanormal} that there is no edge between $U_i$ and $U_j$ if and only if $\Theta_{ij} = 0$. The nonparanormal estimator uses a two-step procedure: (1) for each variable, replace the observations with their normal scores, subject to a Winsorized truncation, and (2) apply the graphical lasso to the transformed data to estimate $\Theta$.

Once we obtain an estimated sparse precision matrix $\widehat{\Theta}$ via the graphical lasso or nonparanormal estimator, we can determine the corresponding graph structure by identifying the zero-pattern of $\widehat{\Theta}$. Independent blocks of random variables (i.e. utterances) can thus be formed by distinct connected components of the estimated graph. To be more specific, let $\{\widehat{S}_k\}_{k=1}^K$ be the set of indices for $K$ connected components in the estimated graph, then under the Gaussian or nonparanormal assumption, we have $U_i \indep U_j$ for any $i\in \widehat{S}_{k'}$ and $j\in \widehat{S}_{k''}$ where $i\neq j$ and $k'\neq k''$. Here, it is worth noting that uncorrelated and joint normality implies independent random variables. Although in this paper we only utilize the inferred blocks in significance analysis for WER, the estimated covariance matrix $\widehat{\Theta}^{-1}$ from the method could be useful in other practical applications.

\subsection{Bootstrap with Inferred Blocks}
Once the set of independent blocks $\{\widehat{S}_k\}_{k=1}^K$ for utterances is estimated, we can apply the blockwise bootstrap \cite{liu2020statistical} to compute the 95\% confidence interval for any WER related statistic of interest.

For any $b=1,\ldots,B^{boot}$ where $B^{boot}$ is a large number, we randomly sample with replacement $K$ elements $\widehat{S}_{1}^{(b)}, \ldots, \widehat{S}_{K}^{(b)}$ from the set of $\{\widehat{S}_k\}_{k=1}^K$ to generate a bootstrap sample
\begin{equation}
\{(m_i, e_i^A, e_i^B)\}_{i\in \widehat{S}_{k'}^{(b)}}, k'=1,\ldots,K,
\end{equation}
where each $\widehat{S}_{k'}^{(b)}\in\{\widehat{S}_k\}_{k=1}^K$. Then for this bootstrap sample, the statistics for WER of ASR system \emph{A},  absolute and relative WER difference of system \emph{B} versus system \emph{A}, can be computed as
\begin{align}
    W_{A}^{(b)}&=\frac{\sum_{k'=1}^K\sum_{i\in \widehat{S}_{k'}^{(b)}}e_i^A}{\sum_{k'=1}^K\sum_{i\in \widehat{S}_{k'}^{(b)}}m_i}, \\
    \Delta W_{abs}^{(b)}&=\frac{\sum_{k'=1}^K\sum_{i\in \widehat{S}_{k'}^{(b)}}(e_i^B-e_i^A)}{\sum_{k'=1}^K\sum_{i\in \widehat{S}_{k'}^{(b)}}m_i}, \\
    \Delta W_{rel}^{(b)}&=\frac{\sum_{k'=1}^K\sum_{i\in \widehat{S}_{k'}^{(b)}}(e_i^B-e_i^A)}{\sum_{k'=1}^K\sum_{i\in \widehat{S}_{k'}^{(b)}}e_i^{A}},
\end{align}
respectively. Once we have all $\{\Delta W_{rel}^{(b)}\}_{b=1}^{B^{boot}}$ computed, the 95\% confidence interval for the statistic $\Delta W_{rel}$ can be calculated by their empirical percentiles at 2.5\% and 97.5\%:
\begin{equation}
\label{per}
\big[\text{percentile}(\Delta W_{rel}^{b}, 2.5\%), \text{percentile}(\Delta W_{rel}^{b}, 97.5\%)\big].
\end{equation}
Here, a 95\% confidence interval means that if we were able to have 100 different datasets from the same distribution of original dataset, and compute a 95\% confidence interval based on each of these 100 datasets, then around 95 out of these 100 confidence intervals will contain the true value of the statistic of interest \cite{neyman1937x, stuart1963advanced, cox1979theoretical}. 

The confidence intervals for $W_{A}$ or $\Delta W_{abs}$ can be calculated in a similar manner with above.

\section{Theoretical Properties}
\label{properties}
We present theories to show the blockwise variance estimator with inferred blocks from the graphical lasso is statistically consistent for WER estimation under some mild conditions. Specifically, as the number of evaluation utterances increases indefinitely, the resulting sequence of variance estimates converges to the truth variance \cite{amemiya1985advanced}.

For simplicity, suppose all utterances in the evaluation set have the same number of words in the reference, that is, $m_i=m$ for all $i=1,\ldots,n$. Denote $Z_i:=e_i/m$. Then the statistic of interest can be written as
\begin{equation}
    W_n = \frac{1}{n}\sum_{i=1}^n Z_i.
\end{equation}
Further, assume the $n$ utterances can be divided into $K_n^*$ nonoverlapping and independent blocks, denoted as $\{S_k\}_{k=1}^{K_n^*}$, with each block having the same number of utterances as $d_n^*$.

Suppose $\{\widehat{S}_k\}_{k=1}^{K_n}$ is the set of estimated independent blocks from the graphical lasso. Following Theorem 1 in \cite{raskutti2008model}, under some conditions on the precision matrix of Gaussian distribution and the scaling of triple $(L, D, n)$, the set of $\{\widehat{S}_k\}_{k=1}^{K_n}$ approaches the true set of $\{S_k\}_{k=1}^{K_n^*}$ with a high probability when the embedding dimension $L$ is large enough. Here, $D$ represents the maximum of node degrees in graph $G$. Specifically
\begin{equation}
\label{conv}
\mathbb{P}\big(\{\widehat{S}_k\}_{k=1}^{K_n}=\{S_k\}_{k=1}^{K_n^*}\big)\geq 1 - \exp(-c\log n)\rightarrow 1,
\end{equation}
for some constant $c>0$.

We further let
\begin{equation}
    W_{\widehat{S}_k, n}:=\frac{1}{|\widehat{S}_k|}\sum_{i\in\widehat{S}_k}Z_i
\end{equation}
be the statistic of interest from the $k$th inferred block, where $|\widehat{S}_k|$ refers to the cardinality of the set of $\widehat{S}_k$. Consider the corresponding blockwise variance estimator
\begin{equation}
    \hat\sigma_n^2:=\frac{d_n}{K_n}\sum_{k=1}^{K_n}{\left(W_{\widehat{S}_k, n}- W_n\right)}^2,
\end{equation}
where $d_n=[n/K_n]$ denotes the average number of utterances in the inferred blocks. The following establishes its $L_2$-consistency.

\begin{theorem}
\label{thm:main}
Assume the asymptotic variance of $W_n$ is 
\begin{equation}
    \lim_{n\rightarrow\infty}n\mathbf{E}(W_n-\mathbf{E}( W_n))^2=\sigma^2\in(0,\infty)
\end{equation}
and $\mu=\mathbf{E}(Z_i)$ for any $i=1,\ldots,n$. Let $d_n^*$ be s.t. $d_n^*\rightarrow\infty$ and $K_n^*\rightarrow\infty$ as $n\rightarrow\infty$. If $n^2\mathbf{E}(W_n-\mu)^4$ is uniformly bounded and the conditions of Theorem 1 in \emph{\cite{raskutti2008model}} are satisfied, then
\begin{equation}
    \hat\sigma_n^2\rightarrow_{L_2}\sigma^2\;as\;n\rightarrow\infty.
\end{equation}
\end{theorem}

The proof is straightforward by connecting the graph structure recovery property (\ref{conv}) with Theorem 1 in \cite{liu2020statistical}. Specifically, the integrals of expectation and variance of $\hat\sigma_n^2$ can be decomposed into the correct graph structure recovery part (with a high probability) and the incorrect graph structure recovery part (with a low probability). The former follows the proof of Theorem 1 in \cite{liu2020statistical}. Notice that each $(W_{\widehat{S}_k, n}- W_n)^2$ is bounded at the worse case, then the latter part can be dominated by $\exp(-c\log n)$ and become negligible as $n$ goes to infinity.

This theorem shows statistical guarantees in the dependent structure inference for utterances and provides the convergence property of the corresponding blockwise estimator.

\section{Experiments}
\label{experiments}

\subsection{Dataset and Setup}
We experiment with the {LibriSpeech} data \cite{panayotov2015librispeech}, consisting of 960 hours transcribed training utterances from 2,338 speakers. The evaluation dataset has the splits of \texttt{test-clean} and \texttt{test-other}.

Table~\ref{tab:data} shows detail of {LibriSpeech} evaluation set on number of utterances and number of speakers.

\begin{table}[ht!]
 \caption{Summary of {LibriSpeech} evaluation dataset.}
  \centering
  \begin{tabular}{l|c|c}
    \toprule
    \emph{} & \texttt{test-clean} & \texttt{test-other} \\
    \midrule
    {Number of utterances} & 2,620 & 2,939\\
    {Number of speakers} & 40 & 33 \\
    \bottomrule
  \end{tabular}
  \label{tab:data}
\end{table}
We consider two ASR systems \emph{A} and \emph{B} in this investigation. The ASR system \emph{A} is a RNN-T model with the Emformer encoder \cite{emformer2021streaming}, LSTM predictor, and a joiner. The ASR system \emph{B} utilizes the fast-slow cascaded encoders \cite{mahadeokar2022streaming}. Both models have around 80 million parameters in total and are trained from scratch using the training utterances of {LibriSpeech}.

Table~\ref{tab:wer} displays the empirical WERs of both ASR systems as well as the absolute and relative WER difference of ASR system \emph{B} compared with system \emph{A}. The reported WERs are multiplied by 100.0 as per convention. We observe that ASR system \emph{B} achieves large improvement over system \emph{A}.

\begin{table}[ht!]
 \caption{WER results of ASR systems \emph{A} and \emph{B}.}
  \centering
  \begin{tabular}{l|c|c}
    \toprule   
    \emph{} & \texttt{test-clean} & \texttt{test-other} \\
    \midrule
    {WER of ASR System \emph{A}} & 3.77 & 10.10 \\
    {WER of ASR System \emph{B}} & 3.25 & 8.13 \\
    \midrule
    {Absolute WER difference} & -0.52\% & -1.97\% \\
    {Relative WER difference} & -13.8\% & -19.5\% \\
    \bottomrule
  \end{tabular}
  \label{tab:wer}
\end{table}

With this setup, the goal is to calculate the 95\% confidence intervals for WER of ASR system \emph{A}, relative and absolute WER difference between ASR system \emph{B} and system \emph{A}. We compare various approaches in the experiments including vanilla bootstrap (\textsf{Bootstrap}) \cite{bisani2004bootstrap}, blockwise bootstrap (\textsf{BlockBootstrap}) \cite{liu2020statistical}, and the proposed method (\textsf{InferredBlockBootstrap}).

We set $B^{boot}$ to 10,000 for each bootstrap based method. For the \textsf{BlockBootstrap} method, we treat the utterances from the same speaker as a block. To apply the \textsf{InferredBlockBootstrap} approach, the pre-trained BERT (base model, uncased) \cite{devlin2018bert, wolf2020transformers} is leveraged to obtain the sentence embedding of each reference in evaluation utterances. The representation is in dimension of 768. Then among the utterances of each speaker, we adopt the graphical lasso on their sentence embedding vectors to estimate the underlying graph structure as well as corresponding independent blocks. Cross-validation is utilized to select the regularization parameter $\lambda$. Then the blockwise bootstrap method is applied on the inferred blocks. Here, the assumption we made is that ASR errors on any utterances from different speakers are independent with each other, while any utterances from the same speaker but having very different semantic or linguistic characteristics are also considered as independent with each other.

\subsection{Results}
For each method in the comparison, Table~\ref{tab:ci_a} shows the calculated 95\% confidence intervals for WER of ASR system \emph{A}. For both of \texttt{test-clean} and \texttt{test-other} sets, we can see that the 95\% confidence intervals from \textsf{InferredBlockBootstrap} are wider than the ones generated from \textsf{Bootstrap}, but are narrower than the ones computed by \textsf{BlockBootstrap}.

\begin{table}[ht!]
 \caption{Confidence intervals for WER of ASR system \emph{A}.}
 \centering
  \resizebox{0.9\columnwidth}{!}{%
  \begin{tabular}{l|c|c}
    \toprule  
    \emph{} & \texttt{test-clean} & \texttt{test-other} \\
    \midrule
    {\textsf{Bootstrap}} & [3.54, 4.00] & [9.66, 10.53] \\
    {\textsf{BlockBootstrap}} & [3.33, 4.23] & [8.40, 12.20] \\
    {\textsf{InferredBlockBootstrap}} & [3.41, 4.14] & [8.79, 11.65] \\   
    \bottomrule
  \end{tabular}
  \label{tab:ci_a}
  }
  \vspace{0.3cm}
\end{table}

Table~\ref{tab:ci_abs} and Table~\ref{tab:ci_rel} also display the computed 95\% confidence intervals for the absolute and relative WER difference between ASR systems \emph{B} and \emph{A}, respectively. The findings are similar to the ones that we observe above. On the \texttt{test-other} dataset in Table~\ref{tab:ci_rel}, the width of the 95\% confidence interval from \textsf{Bootstrap} is 4.8\%, while the width of the one from \textsf{BlockBootstrap} is 7.9\%. Also the 95\% confidence interval from \textsf{InferredBlockBootstrap} has the width of 6.7\%, which is around 40\% wider than the one from \textsf{Bootstrap} but 15\% narrower than the one from \textsf{BlockBootstrap}.

We observe that all of these 95\% confidence intervals computed in Table~\ref{tab:ci_abs} and Table~\ref{tab:ci_rel} do not cover the point of 0, which means that the ASR system \emph{B} is statistically significantly better than the system \emph{A} in term of ASR quality. 

Seen from these results, the vanilla \textsf{Bootstrap} method ignores the dependency among utterances and thus leads to narrower confidence intervals, which might result in over-optimistic conclusions due to false-positive discoveries. At another extreme, due to the subjective definitions of block structures, the \textsf{BlockBootstrap} method could be conservative on claiming statistical significance, where the corresponding confidence intervals are wider in the experiments. Instead, the presented \textsf{InferredBlockBootstrap} approach utilizes data-driven methods to explicitly model the dependent structure among utterances and draw sound conclusions. 

\begin{table}[ht!]
 \caption{Confidence intervals for the absolute WER difference of ASR system \emph{B} compared with ASR system \emph{A}.}
  \centering
  \resizebox{\columnwidth}{!}{%
  \begin{tabular}{l|c|c}
    \toprule  
    \emph{} & \texttt{test-clean} & \texttt{test-other} \\
    \midrule
    {\textsf{Bootstrap}} & [-0.67\%, -0.35\%] & [-2.24\%, -1.69\%] \\
    {\textsf{BlockBootstrap}} & [-0.71\%, -0.32\%] & [-2.63\%, -1.44\%] \\
    {\textsf{InferredBlockBootstrap}} & [-0.70\%, -0.34\%] & [-2.50\%, -1.53\%] \\   
    \bottomrule
  \end{tabular}
  }
  \label{tab:ci_abs}
  \vspace{0.3cm}
\end{table}

\begin{table}[ht!]
 \caption{Confidence intervals for the relative WER difference of ASR system \emph{B} compared with ASR system \emph{A}.}
 \centering
 \resizebox{\columnwidth}{!}{%
  \begin{tabular}{l|c|c}
    \toprule  
    \emph{} & \texttt{test-clean} & \texttt{test-other} \\
    \midrule
    {\textsf{Bootstrap}} & [-17.4\%, -9.6\%] & [-21.9\%, -17.1\%] \\
    {\textsf{BlockBootstrap}} & [-18.0\%, -8.9\%] & [-23.5\%, -15.6\%] \\
    {\textsf{InferredBlockBootstrap}} & [-18.1\%, -9.2\%] & [-22.8\%, -16.1\%] \\   
    \bottomrule
  \end{tabular}
  }
  \label{tab:ci_rel}
\end{table}

\begin{figure}[ht!]
  \begin{minipage}[c]{0.494\linewidth}
   \centering
    \includegraphics[width=\linewidth]{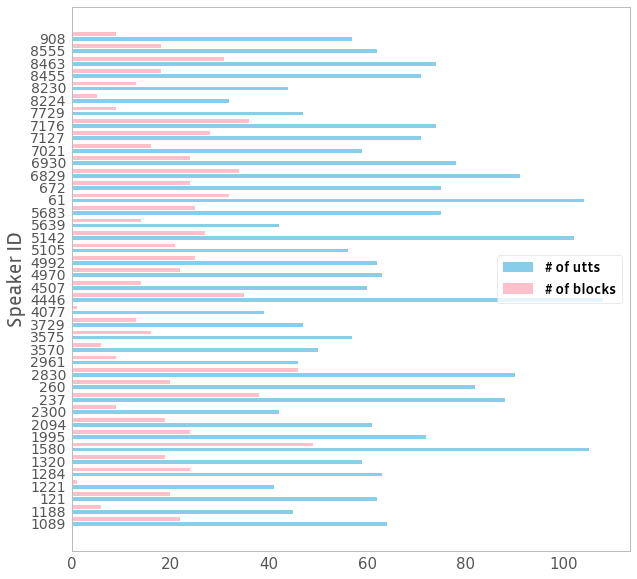} \\
    (a) \texttt{test-clean}
    \end{minipage}
  \hfill
  \begin{minipage}[c]{0.494\linewidth}
    \centering
    \includegraphics[width=\linewidth]{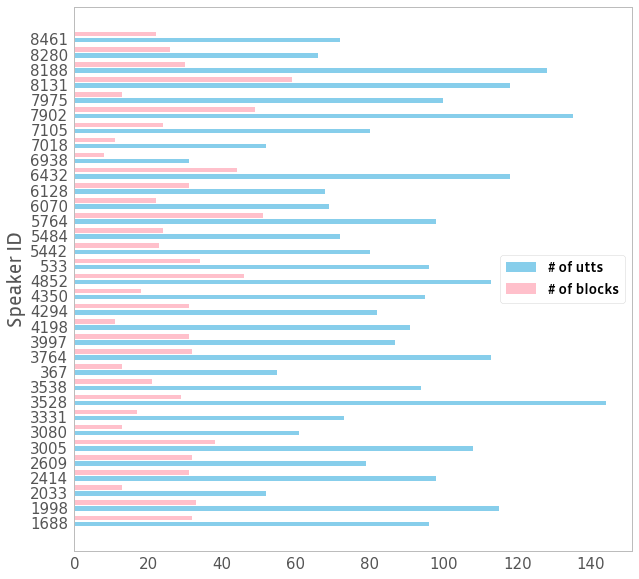} \\
    (b) \texttt{test-other}
  \end{minipage}
  \caption{Number of utterances and number of inferred blocks for each speaker.}
  \label{fig:bar}
\end{figure}

Figure~\ref{fig:bar} shows the number of utterances and number of inferred blocks from the \textsf{InferredBlockBootstrap} method for each speaker. In total, there exist 822 blocks for the \texttt{test-clean} set and 912 blocks for the \texttt{test-other} set. The median of the ratios between the block count and utterance count per speaker is around 0.30.

For the introduced \textsf{InferredBlockBootstrap} approach, we also experiment with the use of nonparanormal, which weakens the Gaussian assumption imposed on the embedding vectors for evaluation utterances. Table~\ref{tab:ci_a_nonpara} shows the computed 95\% confidence intervals on WER of ASR system \emph{A} for graphical lasso and nonparanormal based \textsf{InferredBlockBootstrap}. From the results, we observe that nonparanormal based method results in wider confidence internals compared with the ones obtained from graphical lasso, but they are still narrower than those computed by \textsf{BlockBootstrap}. 

\begin{table}[ht!]
 \caption{Comparison of graphical lasso and nonparanormal based \textsf{InferredBlockBootstrap} on the computation of confidence intervals for WER of ASR system \emph{A}.}
 \centering
  \resizebox{0.9\columnwidth}{!}{%
  \begin{tabular}{l|c|c}
    \toprule  
    \emph{} & \texttt{test-clean} & \texttt{test-other} \\
    \midrule
    \textsf{InferredBlockBootstrap} & & \\
    {$\boldsymbol{\cdot}$ \textsf{graphical lasso}} & [3.41, 4.14] & [8.79, 11.65] \\   
    {$\boldsymbol{\cdot}$ \textsf{nonparanormal}} & [3.37, 4.18] & [8.47, 12.07] \\ 
    \bottomrule
  \end{tabular}
  \label{tab:ci_a_nonpara}
  }
\end{table}

\section{Conclusions}
\label{conclusion}
In this paper we propose to use the graphical lasso and its variant of nonparanormal to estimate the dependent structure among speech utterances in ASR evaluation. We demonstrate its applications on WER significance analysis through blockwise bootstrap resampling with inferred independent blocks. Results are also presented on the {LibriSpeech} dataset for the comparison of 95\% confidence intervals computed from various approaches.



\bibliographystyle{IEEEbib}
\bibliography{refs}

\end{document}